\documentclass[aps,prl,preprint,superscriptaddress, showpacs]{revtex4-1}

\usepackage{graphicx}
\usepackage{dcolumn}
\usepackage{bm}
\usepackage{amsmath}
\usepackage{amssymb}
\usepackage{color}
\usepackage{natbib}
\begin{document}


\title{Evaluation of Doppler shifts in $d$-wave superconductor tunneling junctions}



\author{Takashi Sakamori}
\affiliation{Department of Applied Physics, Nagoya University, Nagoya 464-8603, Japan}

\author{Kenki Matsuoka}
\affiliation{Department of Applied Physics, Nagoya University, Nagoya 464-8603, Japan}

\author{Mitsuhiro Teshigawara}
\affiliation{Department of Applied Physics, Nagoya University, Nagoya 464-8603, Japan}

\author{Yasunori Mawatari}
\affiliation{National Institute of Advanced Industrial Science and Technology (AIST), Tsukuba, Ibaraki 305-8568, Japan}

\author{Keiji Yada}
\affiliation{Department of Applied Physics, Nagoya University, Nagoya 464-8603, Japan}

\author{Yukio Tanaka}
\affiliation{Department of Applied Physics, Nagoya University, Nagoya 464-8603, Japan}

\author{Satoshi Kashiwaya}
\affiliation{Department of Applied Physics, Nagoya University, Nagoya 464-8603, Japan}


\date{\today}

\begin{abstract}
The energy levels of quasiparticles in superconductors experience Doppler shifts due to the influence of a finite current flow. 
The influence of the Doppler shift is particularly pronounced in unconventional superconductors with gap nodes. 
In the case of a normal/insulator/superconductor junction of $d$-wave superconductors, a zero-bias conductance peak exhibits a peak shift in an applied magnetic field. 
Although the shifts have been detected in experiments, the shifts showed different field responses depending on each experiment. 
In this paper, we evaluate the magnitude of the shift based on realistic junction shapes to elucidate the origin of the variability in experimental Doppler shifts. 
We find that the peak shift depends on the junction shape; the splitting is significantly suppressed when the junction is formed at a protruding position. 
Also, the shift is suppressed by the penetration of magnetic flux quanta, resulting in a nonlinear response to the magnetic field. 
The present results explain the origin of the variability in experimental Doppler shifts.
\end{abstract}

\pacs{}

\maketitle

\section{Introduction}
The Meissner effect is one of the fundamental phenomena of superconductivity in which a magnetic field is expelled from superconductors. 
The field is not entirely eliminated from a superconductor; a finite magnetic field resides within a region of approximately the magnetic penetration depth from the surface. 
The magnetic flux can also penetrate superconductors as the unit of flux quanta in type II superconductors. 
When a finite field is present within superconductors, a shielding current is induced due to the finite vector potential, leading to the energy shift of quasi-particles. 
This effect is known as the Doppler shift\cite{volovik1993} and is known to cause a modulation in the density of quasi-particle states within the superconductor. 
The amplitude of the Doppler shift is considerably smaller than the superconducting energy gap when the magnetic field is sufficiently weaker than the critical field ($H_{C2}$). 
Therefore, the Doppler shift rarely manifests as a dominant effect in conventional superconductors without nodes of the pair potentials.
\par
On the other hand, the Doppler shift has been detected as the prominent effect in unconventional nodal superconductors, mainly in two cases. 
One is the conductance spectra in normal/insulator/superconductor (N/I/S) junctions of $d$-wave and $p$-wave symmetries. 
The conductance exhibits sharp zero-bias conductance peaks (ZBCPs) depending on the surface orientation, reflecting the sign change of the pair potential\cite{kashiwaya1995, tanaka1995, kashiwaya2000}. 
The ZBCP is an ideal probe for detecting small energy shifts since the sharp single peak is formed at zero energy, which is significantly different from the bulk state.
In fact, the appearances of the ZBCP and their splittings in the applied field have been widely detected in N/I/S junction experiments on cuprates
\cite{kashiwaya1995, covington1997, ekin1997, alff1998, wei1998, aprili1999, dagan2000, dagan2000110, igu2000, dagan2001, biswas2002,aubin2002, sharoni2002, miyake2003, greene2003, kashiwaya2004, shan2005, elhalel2007, ngai2010, koren2011, kashiwaya2011}. 
The other is the magnetic field response of specific heat, referred to as the Volovik effect\cite{moler1997,wang2011}. 
The specific heat in an applied field has been applied to identify the pairing symmetry of unconventional superconductors because the density of states at zero energy responds sensitively to the field. 
Here, we focus on the magnetic responses of the ZBCP.
\par
The amplitude of the Doppler shift could have prominent spatial dependence when the junction has a nonflat interface. 
Unlike specific heat, tunneling spectroscopy is sensitive to surface states, which means that tunneling conductance spectra strongly depend on the shape of the interface. 
However, previous theoretical analyses assumed infinite flat interfaces between diamagnetic superconductors and normal metals (we refer \cite{fogelstrom1997} as FRS theory in the following).
Although the influences of the Doppler shift have been theoretically analyzed in various situations \cite{tanaka2002, tanaka-Ito2002, tanaka2002b, tanuma2002, tanaka2009, kashiwaya2014, Burset2015}, the shape dependencies of the superconducting electrodes and the influences of magnetic flux quanta have not been considered. 
On the other hand, previous experimental results on the peak splitting by the Doppler shift performed on cuprates report non-linear responses\cite{covington1997,dagan2000110} as well as the distributed amplitudes of the peak shift, which manifests the presence of unrevealed effects in the tunneling effects.
Therefore, we investigate the influences of junction shape and magnetic flux quantum penetration on the tunneling conductance spectra of N/I/S junctions to clarify the origin of distributed experimental results of the Doppler shift. 
We assume realistic junction shapes with corrugations and calculate the shielding current distribution based on the London equation and the time-dependent Ginzburg-Landau (TDGL) equation by using a finite element method (FEM). 
We evaluate the amplitude of the Doppler shift based on the conductance formula for $d$-wave superconductors, which considers the spatial dependence of the shielding current at the superconductor side.

\section{Calculation method}
\begin{figure*}[htbp]
  \begin{center}
		\includegraphics[width=0.5\linewidth]{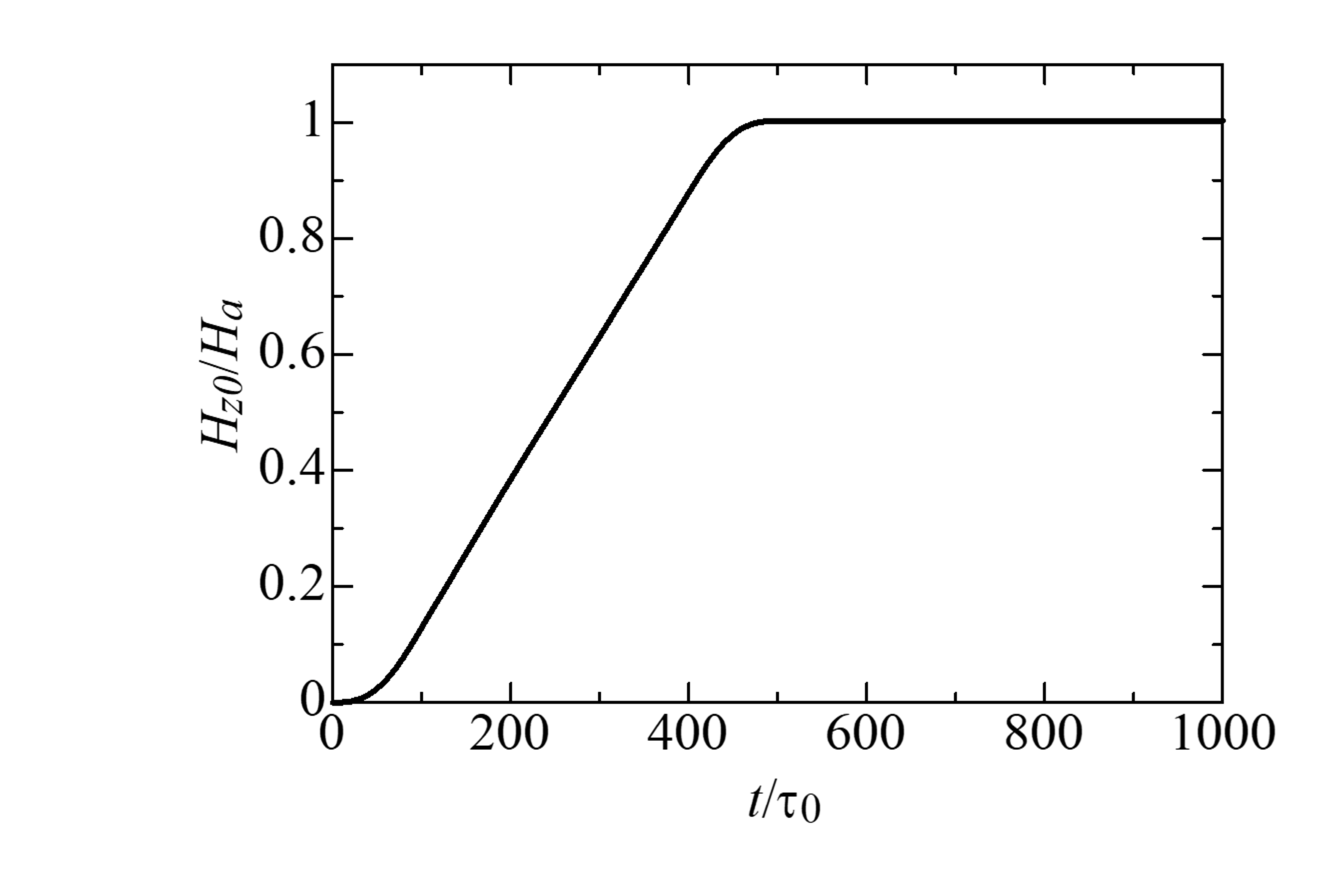}
  \end{center}
    \caption{Time sequence of the magnetic field $H_{z0}$ at the sphere far outside of the superconductor used in the TDGL calculation. 
    At the initial state, the magnetic field $H_z0$ is set to zero, then $H_z0$ is swept to the applied magnetic field $H_a$ within a period of $500\tau_0$ and then kept at  $H_a$ for 500$\tau_0$.}
      \label{Fig1}
\end{figure*}

The calculation method is presented in this section.
The distributions of the magnetic field and vector potential are calculated using the finite element method (FEM, COMSOL Multiphysics\cite{Comsol}) based on the London equation and the TDGL equation. 
First, we consider the case where the applied field $H_a$ is less than $H_{C1}$ and no magnetic flux quanta penetrate the superconductor.
The London equation is used to calculate the vector potential $\bm{A} \equiv (A_x, A_y, A_z)$ in the three-dimensional configuration,
    \begin{align}
        \mathrm{rot}^2 \bm{A} = -      
        \left(
\begin{array}{rrr}
  \frac{1}{\lambda_{xy}^2}& 0&0 \\
  0 &  \frac{1}{\lambda_{xy}^2}&0\\
    0 & 0&  \frac{1}{\lambda_z^2}\\
\end{array}
\right)
\bm{A},
    \end{align}
where $\lambda_{xy}$ and $\lambda_{z}$ are the penetration depth in- and out-of-plane, respectively. 
Considering the two--dimensional properties of cuprate superconductors, we set  $\lambda_{z}=15\lambda_{xy}$ throughout this work. 
The magnetic field distribution above $H_{C1}$ is calculated based on the TDGL equation,
   \begin{align}
        \sigma\frac{\partial \bm{A}}{\partial t}-\kappa^2\{\mathrm{Im}(\psi^\ast \nabla \psi) -|\psi|^2\bm{A}\}-\mathrm{rot}^2\bm{A}=0,
    \end{align}
    \begin{align}
        \frac{\partial \psi}{\partial t}+(i\phi+|\psi|^2-1)\psi+(i\nabla+\bm{A})^2\psi=0.
    \end{align}
Here, $\psi$ is the order parameter ($|\psi|=\Delta_0$ in the bulk), 
$\phi$ is the scalar potential, $\tau_0$ is the relaxation time of the order parameter, and $\sigma$ is normal conductivity.
For the calculation, the length, magnetic field, and time $t$ are normalized by the coherence length ($\xi$),  the upper critical magnetic field ($H_{c2}=\Phi_0/2\pi \xi^2$, $\Phi_0$ is the magnetic flux quantum), and the relaxation time, respectively. 
We use a gauge $\phi=-\frac{\kappa^2}{\sigma} \nabla \cdot \bm{A}$ and the GL parameter $\kappa(=\lambda_{xy}/\xi)$ is fixed at 10 considering the cuprate superconductors.
For the TDGL calculation, we perform two-dimensional calculations by assuming translational symmetry in the $z$-direction due to the memory capacity limitation of our workstation,
thus the anisotropy on in-$xy$-plane  and out-of-plane is neglected.
Therefore, the effect of the diamagnetization is not incorporated into the present TDGL calculation. 
For the boundary conditions of the magnetic field, the calculated magnetic field is set to match the applied magnetic field $H_a$ at a sphere sufficiently far (approximately several times the size of the superconductors) away from the superconductor.
To incorporate the arrangement of magnetic flux quanta, we change the magnetic field at the sphere $H_{z0}$ from zero to $H_a$ within a period of $500\tau_0$, and then fixed the magnetic field for a period of 500$\tau_0$ as shown in Fig.~\ref{Fig1}.
The magnetic field distributions at t=1000$\tau_0$ are used to calculate the conductance spectra.
\par
\begin{figure*}[htbp]
  \begin{center}
		\includegraphics[width=0.5\linewidth]{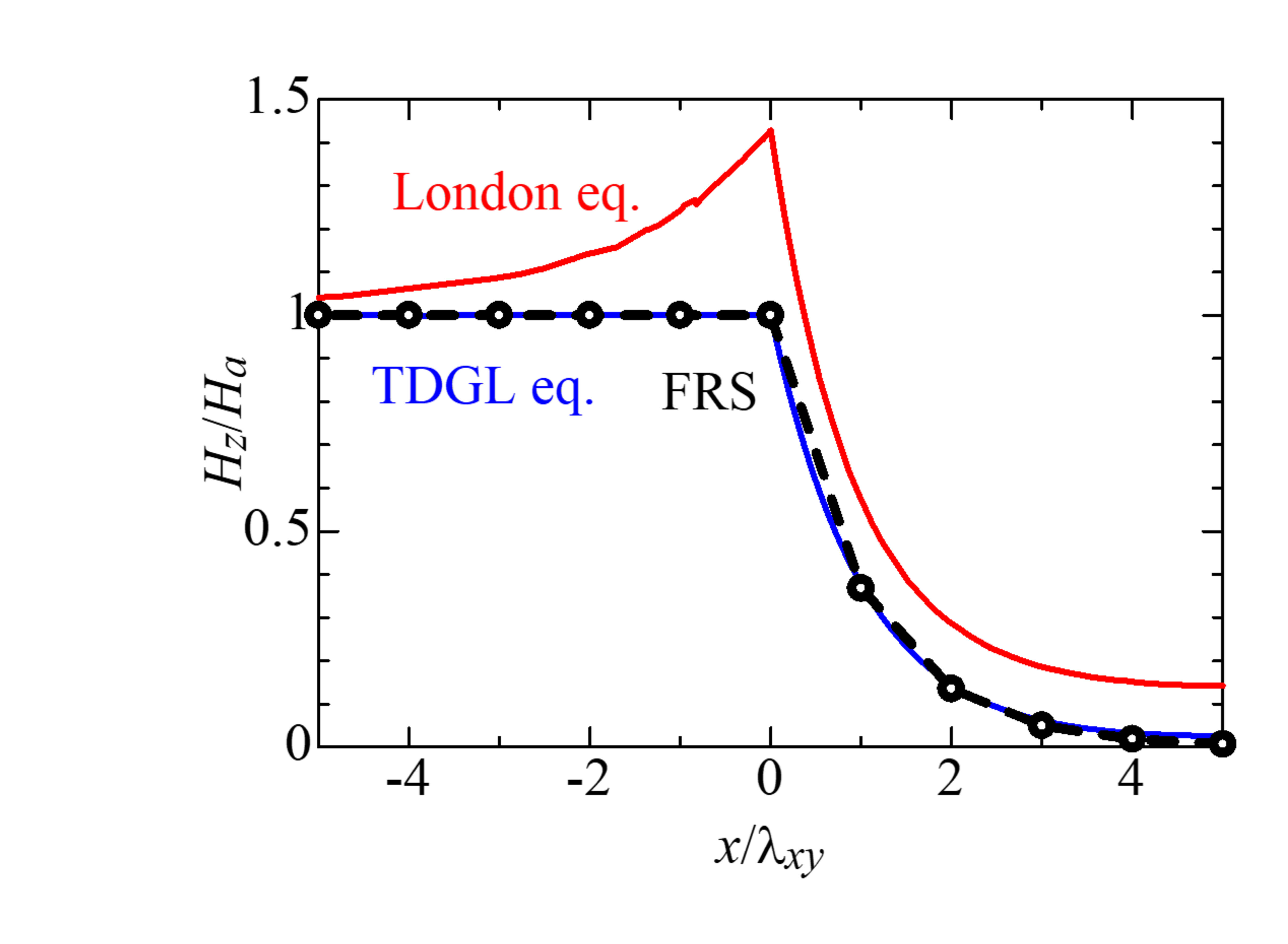}
  \end{center}
  \caption{Magnetic field distributions near the edge of the superconductor in the London equation, TDGL equation, and the FRS theory. 
In the figure, the vertical axis represents the magnetic field for $z$-direction $H_z$ normalized by the applied magnetic field $H_a$, 
and $x<0$ and $x \ge 0$ correspond to the vacuum and superconducting regions, respectively.}
  \label{Fig2}
\end{figure*}

As a result of the London equation calculation in the 3D models, the magnetic field is concentrated and enhanced at the edge of the superconductor due to the diamagnetization effect, as shown in Fig.~\ref{Fig2}.
Correspondingly, the magnitude of the Doppler shift is enhanced, as discussed below. 
On the other hand, the magnetic field obtained by the TDGL equation matches the external magnetic field outside the edges of the superconductor since infinite length in the $z$-direction has been assumed.
We consider the lack of the diamagnetization effect not serious for the latter discussion since this effect is important only for thin films, and tends to be weakened by the penetration of magnetic flux quanta.
In fact, the previous theory~\cite{fogelstrom1997} adopted a boundary condition in which the magnetic field matches the external magnetic field at the edge of the superconductor, as shown in Fig.~\ref{Fig2}.
\par
Based on the calculated magnetic field and vector potential distributions, we evaluate the magnetic field responses of conductance spectra of  N/I/S junctions.
Considering cuprate superconductors, we assume the superconductor has $d_{x^2-y^2}$-wave symmetry in the following analysis.
We use an extended Blonder-Tinkham-Klapwijk (BTK) formalism for $d$-wave superconductors with the interface along [110] orientation (the angle between $x$-axis and $a$-axis of a $d$-wave superconductor is given by $\alpha$=$\pi$/4)\cite{BTK, kashiwaya1995, tanaka1995, kashiwaya2000} by assuming $1/k_F<\xi<\lambda_{xy}$ ($k_F$ is the Fermi wave number).
We consider a situation in which quasiparticles are injected from the normal side with the energy $E$ and angle $\theta$ to the normal of the interface  (see Fig. 3(a)).
The effective pair potentials $\Delta_\pm$ for transmitted quasiparticles to the superconductor are given by $\Delta_{0}\cos(2\theta \mp 2\alpha)$.
The effects of a finite vector potential are introduced following a previous work\cite{tanaka2002}.
The normalized conductance $\sigma_S$ of  is given by
    \begin{align}
        \sigma_S= \frac{\sigma_N}{2}\left(\frac{1+\sigma_N|\Gamma_+|^2+(\sigma_N-1)|\Gamma_+\Gamma_-|^2}{|1
+(\sigma_N-1)\Gamma_+\Gamma_-|^2} 
+\frac{1+\sigma_N|\Gamma_-|^2+(\sigma_N-1)|\Gamma_+\Gamma_-|^2}{|1+(\sigma_N-1)\Gamma_+\Gamma_-|^2}\right),
    \end{align}
    \begin{align*}
        \Gamma_{+}&=\frac{\Delta_{+}^{*}}{E_{+}+\Omega_{+}(E_{+})},  \Gamma_{-}=\frac{\Delta_{-}}{E_{-}+\Omega_{-}(E_{-})}\\
        \Omega_{\pm}(E_{\pm}) &=
        \begin{cases}
            \sqrt{E_{\pm}^2-|\Delta_\pm|^2} & (E>|\Delta_\pm|)\\
            i\sqrt{|\Delta_\pm|^2-E_{\pm}^2} &( -|\Delta_\pm|\leq E\leq |\Delta_\pm|)\\
            -\sqrt{E_{\pm}^2-|\Delta_\pm|^2} & (E<-|\Delta_\pm|)
        \end{cases}\\
		E_{\pm} &=E\mp ev_{F} (A_x\cos\theta \pm A_y\sin\theta).
    \end{align*}
The conductance in the normal state $\sigma_N$ for the $\delta$-function barrier  is given by
    \begin{align}
        \sigma_N = \frac{\cos^2\theta}{Z^2+\cos^2\theta}.
    \end{align}
Here, the barrier parameter $Z$ for the $\delta$-function barrier ($H\dot \delta(x)$, shown in Fig. 3(b)) is described by 
$Z=\frac{mH}{\hbar^2 k_F}$ where $m$ is the electron mass, and $\hbar$ is the Dirac constant.
The barrier parameter for the finite thickness barrier (potential height $U$ with thickness of $d$, shown in Fig. 3(c)) is
    \begin{align}
Z = \frac{\sinh\left( \sqrt{1-\kappa^2 \cos^2\theta}\nu d\right)}{2\kappa\sqrt{1-\kappa^2 \cos^2\theta}} ,
    \end{align}
where $\kappa=\frac{k_F}{\nu}$ and $\nu=\sqrt{\frac{2mU}{\hbar^2}}$.
The integration over all injection angles gives the total conductance of the junction $\sigma_T$,
    \begin{align}
        \sigma_T = \frac{\int_{-\pi/2}^{\pi/2}\sigma_S\cos\theta d\theta}{\int_{-\pi/2}^{\pi/2}\sigma_N\cos\theta d\theta}.
    \end{align}
\begin{figure*}[htbp]
  \begin{center}
		\includegraphics[width=1.0\linewidth]{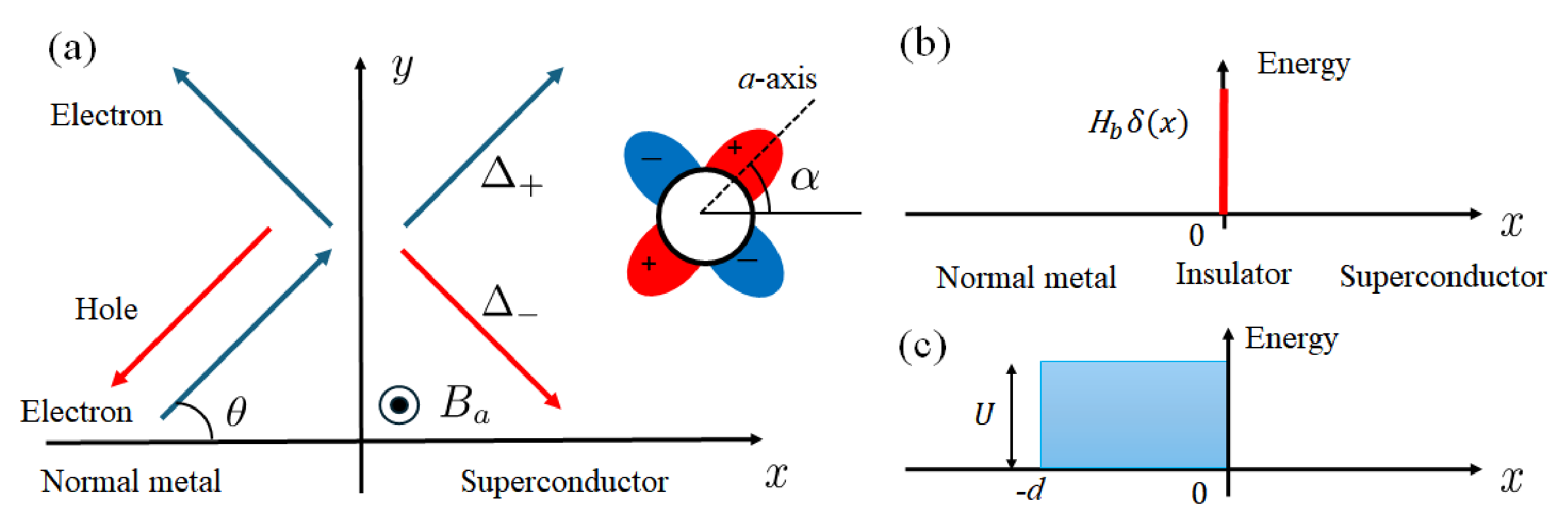}
  \end{center}
  \caption{Schematic diagrams of the N/I/S tunnel junction. 
  (a) We assume the superconductor has $d_{x^2-y^2}$-wave symmetry and the interface ($yz$-plane at $x$=0) coincides with the [110] orientation of the $d_{x^2-y^2}$-wave superconductor.
The angle $\alpha$ between the $a$-axis of and the $x$-axis is fixed at $\pi/4$). 
The$c$-axis of the $d_{x^2-y^2}$-wave superconductor is parallel to $z$-axis. 
Quasiparticles are injected from the normal metal with angle $\theta$ to the interface normal. 
The arrows represent the direction of the group velocity of each quasiparticle. 
(b) Potential profile of the N/I/S junction of the $\delta$-function barrier. 
The conductance spectra except Fig. 12 are calculated with $Z$=5 ($H_b=5\frac{\hbar {k_F}^2}{m}$, $m=m^*/2$).
(c) Potential profile of the finite thickness barrier with the potential height of $U$ and width of $d$.}
\end{figure*}
The characteristic magnetic field at which the energy shift due to the magnetic field becomes $\Delta_0$ is given by $H_0$ (=$\frac{\Phi_{0}}{\pi^2 \xi \lambda_{xy}}$). 
To neglect the reduction in the gap amplitude due to the magnetic field, the magnetic field $H_a$ is assumed to be sufficiently smaller than $H_0$.
The peak shift due to the Doppler shift is evaluated by the energy of the split peak in the conductance $\sigma_T$.
Differently from a previous work\cite{tanaka-Ito2002}, we ignore the anisotropy of the Fermi surface in the $xy$-plane by assuming simple cylindrical shapes for the Fermi surfaces both in the superconductor and normal metal.

\section{Results}
The results of the calculation are presented in this section.
The magnetic field response below $H_{C1}$ in the London equation framework is linear to the field, and the applied field was fixed to 0.1$H_0$.
To clarify the effects of the non-flat interface, superconducting electrode shapes without corrugation (Fig. 4(a)) and without corrugation (Fig. 4(b)) were employed.
Considering a thin film sample, the thickness of superconductors was set to 3$\lambda_{xy}$.
The N/I/S junctions were assumed to be located at the positions marked by A-F in the figure.
The junction areas of $0.3\lambda_{xy} \times 0.3\lambda_{xy}$ size were divided into $10\times10$ equal-sized sub-areas and summed the conductance under the finite vector potential with equal weight. 
 \begin{figure*}[htbp]
  \begin{center}
		\includegraphics[width=1.0\linewidth]{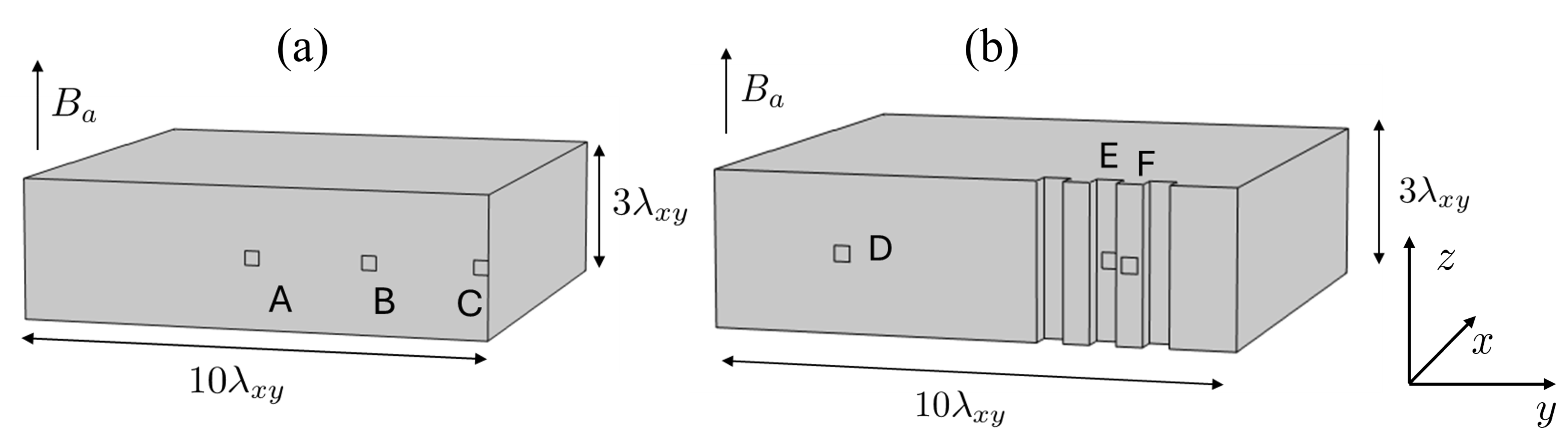}
  \end{center}
  \caption{Schematic diagrams of the superconductor electrode shapes and junction positions. 
  (a) Rectangular without corrugations, (b) rectangular with a dip at an edge. 
  The thickness of the superconductors was set to 3$\lambda_{xy}$ in both cases. 
  The applied magnetic field is parallel to the $z$-axis, and $A$-$F$ represents the position of junctions.}
\end{figure*}
\par
The magnetic field distributions are presented in Fig. 5.
The field is enhanced near the center of the straight edge due to the diamagnetization field effect, while the magnetic field is weakened around the corners.
\begin{figure*}[htbp]
  \begin{center}
		\includegraphics[width=1.0\linewidth]{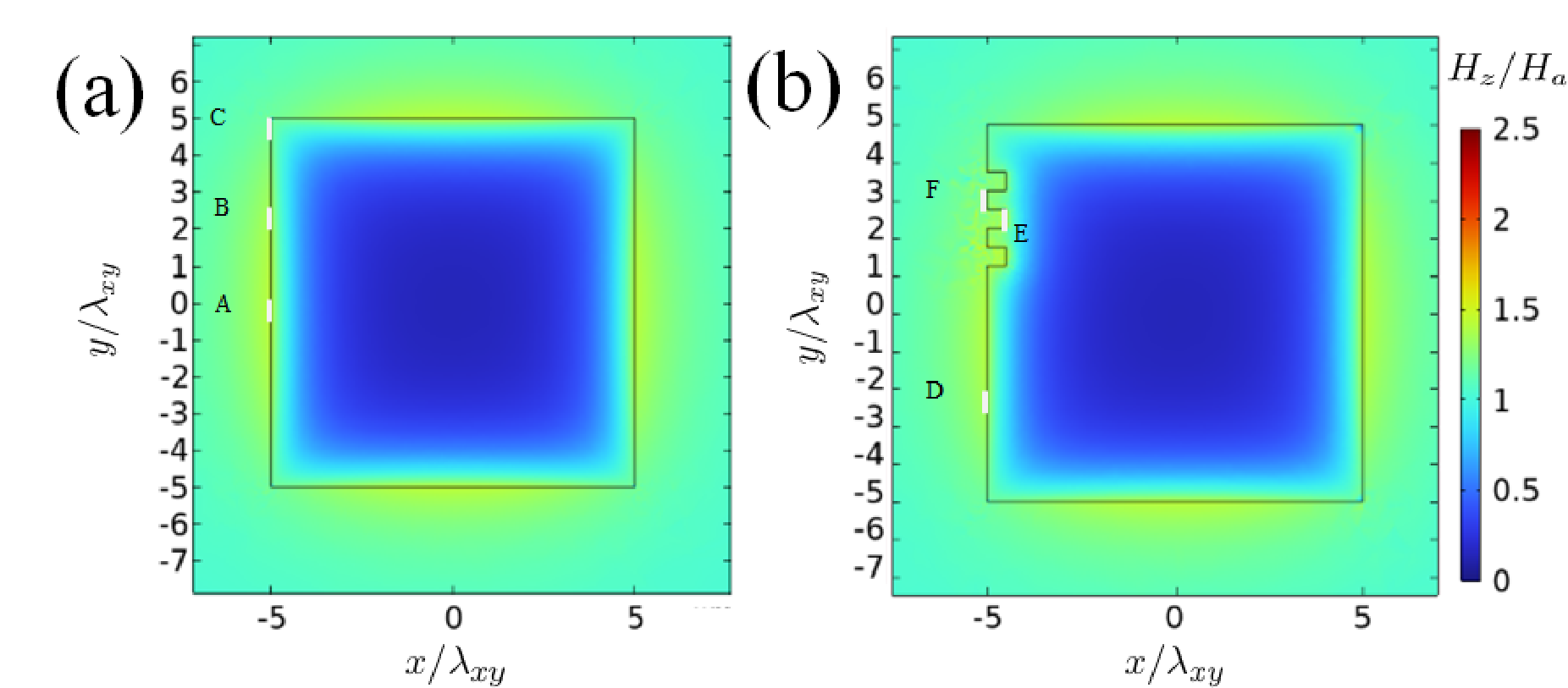}
  \end{center}
  \caption{The spatial distribution of $H_z$ at $H_a=0.1H_0$ calculated by the London quation. (a) Rectangular without corrugations, (b) rectangular with a dip at an edge.}
\end{figure*}
This is reasonable because the shielding current tends to be weakened near the corners and the dip.
To clarify the shielding current distribution, $J_S$ parallel to the interface on the side edge (Fig. 5(a)) is plotted in Fig. 6.
\begin{figure*}[htbp]
  \begin{center}
		\includegraphics[width=0.8\linewidth]{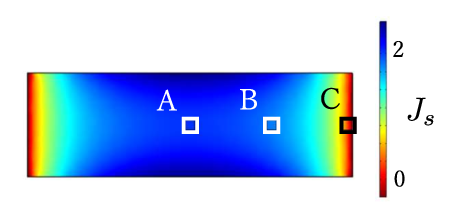}
  \end{center}
  \caption{The shielding current density distribution $J_S$ in the direction parallel to the interface at the edge surface of Fig. 4(a).
  The current amplitude is normalized by $\frac{-H_a}{\mu_0\lambda_{xy}}$, which is the expected value for a semi-infinite superconductor edge surface. 
  The shielding current density exceeds unity near the center due to the diamagnetization field effect and tends to be suppressed as approaching to the corner.
  The marks $A$, $B$, and $C$ represent junction positions.}
  \end{figure*}
Due to the diamagnetization effect in the 3D shape, the shielding current density exceeds unity near the center and is suppressed as approaching the corner.
The conductances $\sigma_T$ at the positions $A$-$C$ are shown in Fig. 7(a).
It is clear that the amplitudes of peak shift correspond to the shielding current density.
The peak shift exceeds the theoretical value by FRS at positions $A$ and $B$, which corresponds to the magnetic field enhancement.
We expect such a trend can be detected in actual experiments if the junctions are formed on thin films.
The amplitude of the peak shift is suppressed as approaching the corner, shown by $C$, corresponding to the $J_s$ distribution in Fig. 6.
\begin{figure*}[htbp]
  \begin{center}
		\includegraphics[width=0.8\linewidth]{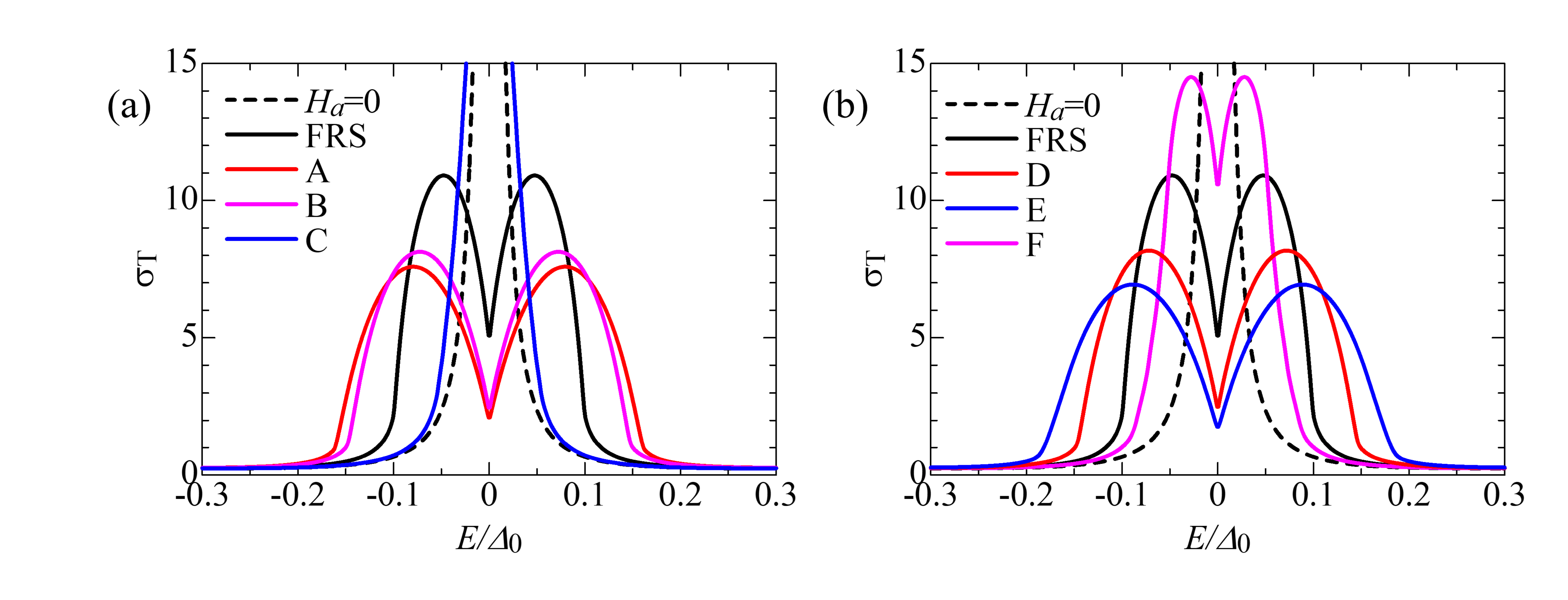}
  \end{center}
  \caption{Calculated conductances $\sigma_T$ in the field of $H_a=0.1H_0$ based on the London equation at the positions (a) $A$-$C$, and  (b) $D$-$F$.} 
\end{figure*}
Next, the effect of corrugations on the surface is discussed.
Figure 5(b) exhibits the magnetic field distribution of the superconductor with corrugations at the surface (Fig. 4(b)) and
Fig. 7(b) shows the conductance $\sigma_T$ calculated at the positions $D$-$F$.
The peak shifts at $D$ and $E$ are almost comparable to those of Fig. 7(a), while the shift is seriously suppressed at protruding position $F$.
The origin of the suppression is similar to that near the corner discussed above, and thus the suppression becomes serious when the slit size is less than $\lambda_{xy}$.
It is important to note that peak shift can be larger or smaller than that based on the FRS theory.
In real experiments, the junction tends to be formed at a protruding position when the junction fabrication relies on the mechanical cleavages of single crystals.
This consideration is consistent with the fact that experimentally detected peak shifts are smaller than the value based on the FRS theory as described below.
\par
The Doppler shifts in the magnetic fields above $H_{C1}$ are analyzed in the framework of the TDGL equation to clarify the effect of magnetic flux quanta penetration.
The electrode shapes used in the calculations are simple squares without or with slits and a dip, as shown in Fig. 8.
The N/I/S junctions were assumed to be formed at the positions $J$-$M$.
To reduce memory consumption during the calculation, we assumed the two-dimensional structure by applying translational invariance for the $z$ direction.
The magnetic field distributions in the superconducting electrodes in $H_a$ = 0.2$H_{C2}$ ($>H_{C1}$) are shown in Fig. 9.
For all cases, the existence of magnetic flux quanta shown by blue spots can be confirmed.
Due to the two-dimensional structure, the magnetic field concentration by the diamagnetization discussed above was not present in this calculation.
Whereas, the shielding current due to the Meissner effect still remains.
Moreover, the amplitude of the peak shift is influenced by the flux quanta arrangement, since the shielding current, whose direction is opposite to that at the edge, exists around the flux quanta.

\begin{figure*}[tb]
  \begin{center}
		\includegraphics[width=0.8\linewidth]{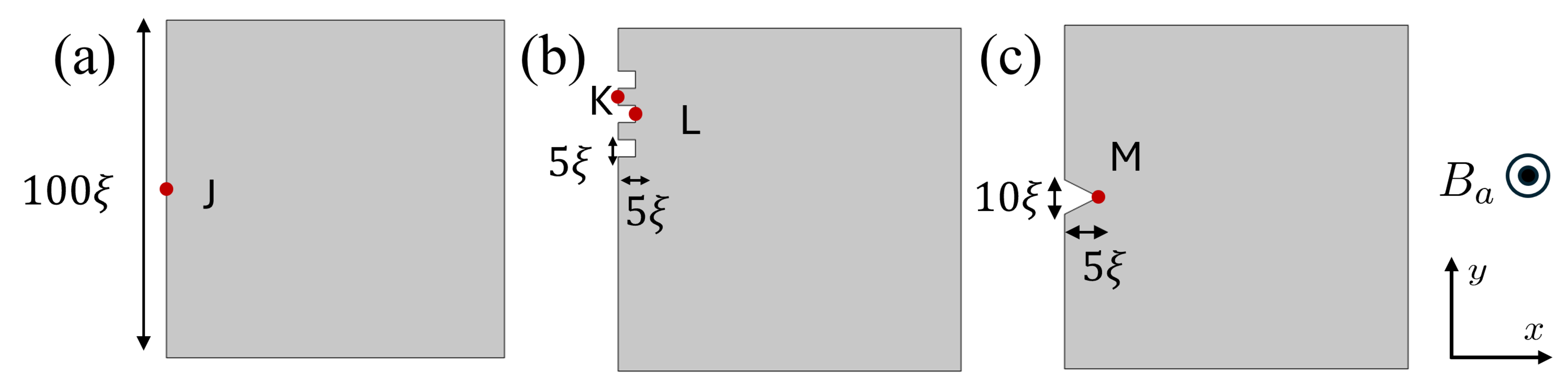}
  \end{center}
  \caption{The shapes of the superconducting electrode used in the TDGL calculation, (A) flat interface, (B) with slits, and (C) with a dip.
  $J$-$M$ represent the positions of the N/I/S junctions. }
\end{figure*}
\begin{figure*}[tb]
  \begin{center}
		\includegraphics[width=0.8\linewidth]{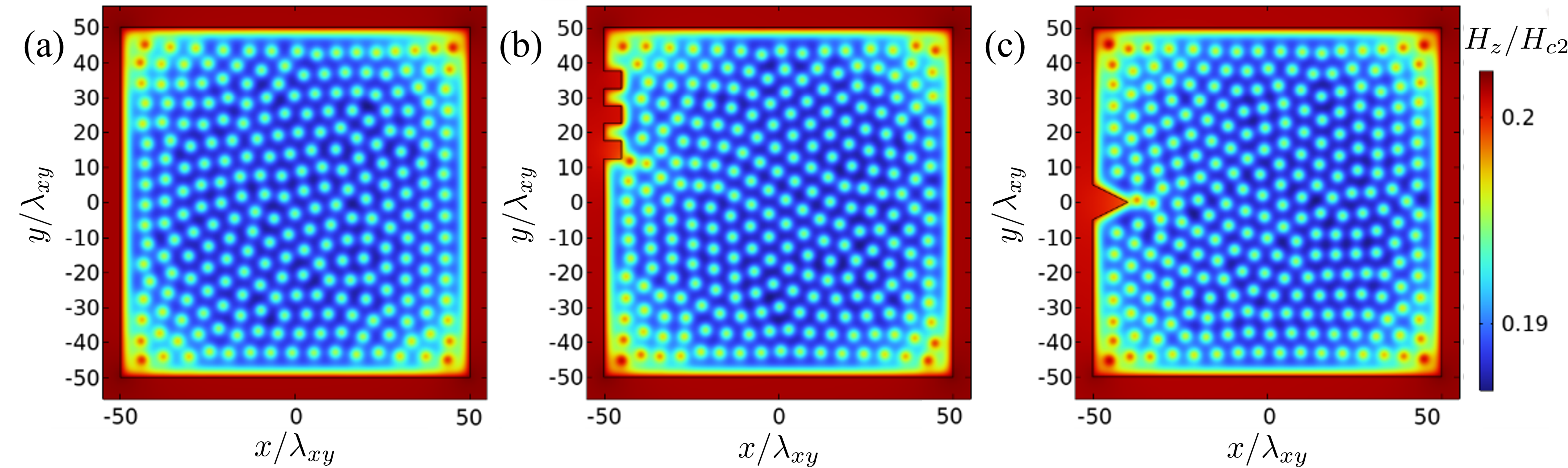}
  \end{center}
  \caption{The magnetic field distribution in the superconducting electrodes calculated by the TDGL equation with the shapes of (A)-(C) shown in Fig. 8.
  The applied field $H_a$ is 0.2$H_{C2}$ ($>H_{C1}$) for all cases. 
  The presence of the shielding current near the edge and around the magnetic flux quanta can be confirmed by the color change.}
\end{figure*}
To further confirm the validity of the calculations, we varied the applied magnetic field in each shape and evaluated the magnetization $M_z$, as shown in Fig. 10.
Here, $M_z$ given by the integration over the plane divided by the area $S$,
    \begin{align*}
         M_Z =\frac{\mu_0}{S} \int (H_z-H_a) dx dy.
    \end{align*}
Evaluated $H_{C1}$s by the magnetic field at which the magnetic flux begins to enter the superconductors are approximately (a) 0.06$H_{C2}$, (b) 0.04$H_{C2}$, and (c) 0.03$H_{C2}$.
The difference of $H_{C1}$ reflects the shape dependence of the surface potential; (a) rectangular shape is highly symmetric and the surface potential blocks the magnetic flux to enter as increasing the applied field.
Whereas the presence of the slits and dip in models (b) and (c) suppresses the surface potential and makes it easier to enter the flux.
After the magnetic flux begins to penetrate, a slow decrease in magnetization is observed, which is consistent with that expected for usual type II superconductors. 
Therefore, we conclude that the present calculation reasonably reproduces the magnetic responses.
\par
Based on the magnetic field distribution, total conductance spectra at the positions $J$-$M$ in Fig. 8 are calculated and the amplitudes of the peak shift are evaluated as shown in Fig. 11.
In this calculation, the junction area is divided into parallelly connected 10 sub-area, and total conductance is calculated by summation of all subareas,  and $A$ in eq. (4) is replaced by $\mu_0 \lambda_{xy}^2 J_s$ to certify the effect of the Doppler shift.
\begin{figure*}[tb]
  \begin{center}
		\includegraphics[width=0.8\linewidth]{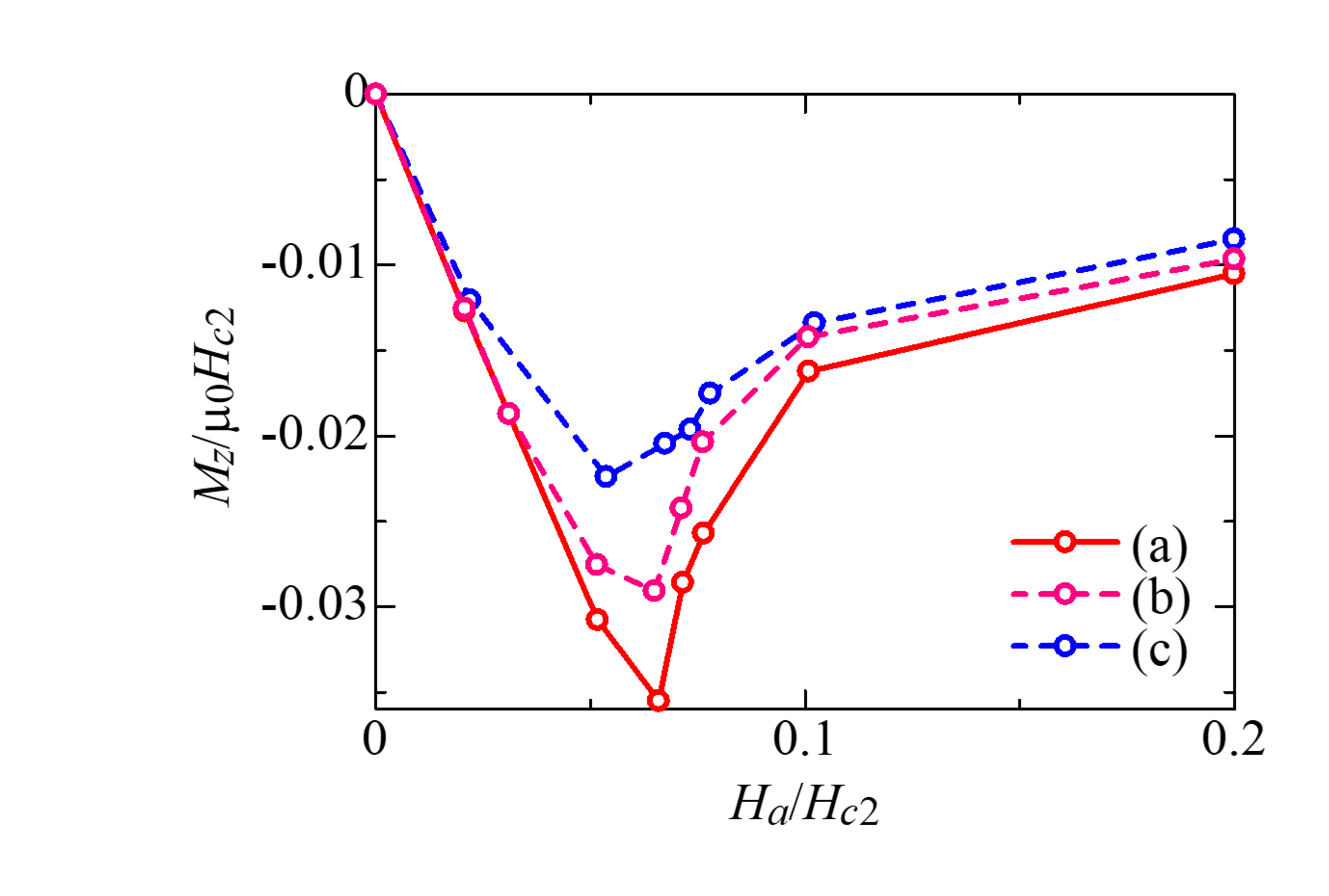}
  \end{center}
  \caption{The magnetizations $M_Z$ of the superconductor electrodes shown in Fig. 8 (a)-(c) based on the TDGL as the function of the applied field. 
  The curves linearly respond to the applied field below approximately 0.02 $H_{C2}$, take the minimum values, and then slowly recover.
  Evaluated  $H_{C1}$s by the magnetic field giving the minimum  $M_Z$ are (a) 0.06$H_{C2}$, (b) 0.04$H_{C2}$, and (c) 0.03$H_{C2}$, respectively.}
\end{figure*}
\begin{figure*}[tb]
  \begin{center}
		\includegraphics[width=0.8\linewidth]{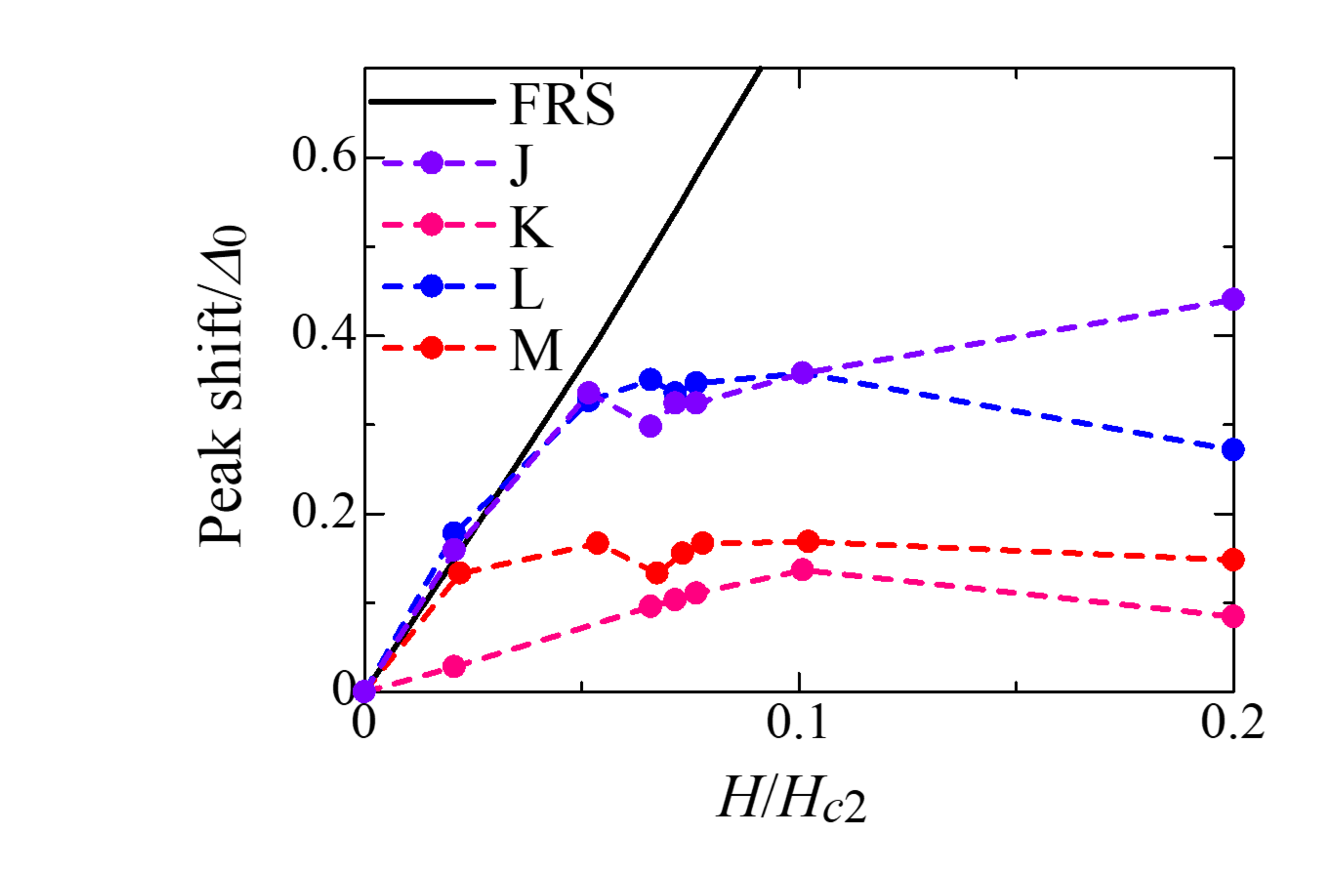}
  \end{center}
  \caption{Evaluated peak shifts due to the Doppler shift in the conductance spectra at junction positions $J$-$M$ in Fig. 8. 
  The peak shift amplitude based on the FRS theory is plotted by the black line.}
  \end{figure*}
We find that the peak shifts are an almost similar level to that of the FRS theory when the magnetic field is smaller than $H_{C1}$.
Whereas they deviate from the FRS by increasing the field.
These non-linear responses are different from those by the London model as well as previous theories\cite{fogelstrom1997,tanaka2002,tanaka-Ito2002}.
Note that the shift begins to decrease by further increasing the field.
This feature reflects that the Doppler shift at the edge is diminished by the circulating current around the magnetic flux when the magnetic flux is located close to the junction position.
Such non-monotonic behavior is consistent with the magnetization shown in Fig. 10.
The maximum shift tends to be suppressed by the presence of the slits and the dip.
By comparing the results of $L$ with $K$, the peak shift is found to be suppressed at the protruding position, similar to the results of the London equation.
This fact clearly indicates that the detailed identification of the superconductor shape and the junction position is important for quantitative analysis of the peak shift of the Doppler shift.

\section{Discussion}
In this section, we compare the above results with experiments. 
Doppler shift has been detected experimentally in YBa$_2$Cu$_3$O$_{7-\delta}$ \cite{covington1997,aprili1999,dagan2001,greene2003,kashiwaya2004,ngai2010} and La$_{1.85}$Sr$_{0.15}$CuO$_4$ \cite{alff1998,miyake2003,yuli2007,teshima2023}.
In these results, the widths of peak splitting were widely dispersed, for example, 0.7 meV/T \cite{covington1997}, 1.2 meV/T \cite{dagan2001}, 0.5 meV/T \cite{aprili1999}, and almost no-splitting \cite{kashiwaya2004} in YBa$_2$Cu$_3$O$_{7-\delta}$, and 0.3 meV/T \cite{alff1998} and 0.06 meV/T \cite{teshima2023} in La$_{1.85}$Sr$_{0.15}$CuO$_4$.
Comparing the above analyses with experimental results, we find three notable features.
At first, the origins of a variety of the Doppler shift have not been clarified in the framework of the FRS theory with the flat interface\cite{fogelstrom1997}.
On the other hand, the present analysis clarifies that the corrugation at the surface seriously modifies the Doppler shift even if the corrugation is small.
Thus the distributed experimental results are shown to be consistent with the responses of the Doppler shift.
Secondly, according to the FRS theory, the amplitude of energy shift is 2.3 meV/T for YBa$_2$Cu$_3$O$_{7-\delta}$ by assuming typical values ($\Delta_0$=17 meV, the coherence length $\xi$=1.4 nm, the penetration depth $\lambda_{xy}$=800 nm), and 1.3meV/T for La$_{1.85}$Sr$_{0.15}$CuO$_4$ ($\Delta_0$=11 mV, $\xi$=3.26 nm, $\lambda_{xy}$=300 nm).
On the other hand, all the experimental results exhibited smaller shifts than those by the FRS theory.
This is quite reasonable because, in real experimental situations, the junction tends to be formed at protruding positions when corrugations exist at the surface.
Thirdly, previously reported nonlinear responses \cite{covington1997,dagan2000110} qualitatively agree with the response expected from the penetration effect of magnetic flux in the TDGL (see Fig. 11). 
Unfortunately, the reentrant behavior in high magnetic fields shown in Fig. 11 has not been detected in experiments thus far.
However, we expect that such behavior could be observed by performing experiments in higher magnetic fields.
\par
Finally, we discuss the effects of the barrier potential shape on the peak shift.
Previous experiments on La$_{1.85}$Sr$_{0.15}$CuO$_4$ \cite{teshima2023} have shown that the magnitude of peak splitting responds to the conditions of the insulating layer formation.
This feature can be understood by a model with a finite width shown in Fig. 3(c).
Figure 13 shows conductance spectra at position $A$ of Fig. 4(a) as the function of the barrier potential shape.
For the calculation of the finite width model, we changed the barrier thickness while keeping the total tunneling probability almost constant.
By comparing the spectrum of $\delta$-function barrier model Fig 12(a), those of the finite potential model (Fig 12(b) and (c)) exhibit the suppression of the peak shift.
The suppression becomes prominent as increasing the barrier thickness.
The origin is explained as follows; the changes in barrier thickness affect the $k$-space distribution (referred to as the tunneling cone) of tunneling electrons. 
Even under conditions where the tunneling probability is fixed, increasing the barrier thickness concentrates the tunneling current on the component perpendicular to the interface (the tunneling cone becomes sharper).
As a result, the peak shift is suppressed by increasing the barrier thickness.
\par
\begin{figure*}[tb]
  \begin{center}
		\includegraphics[width=0.8\linewidth]{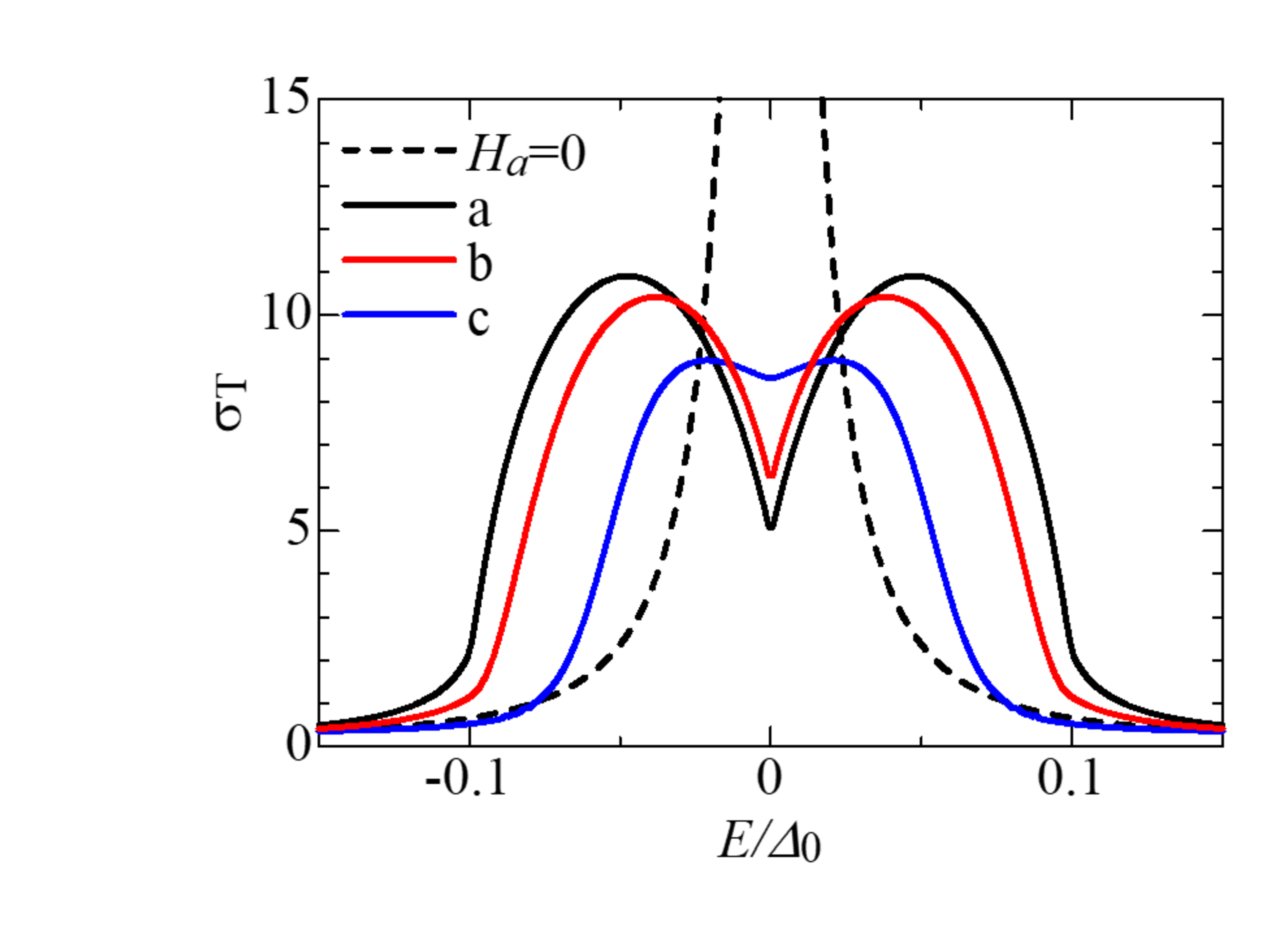}
  \end{center}
  \caption{Barrier potential shape dependence of the Doppler shifts at position $A$ of Fig. 4(a). 
 The total conductance spectra are plotted under the conditions without magnetic field ($H_a=0$, $H_b=5\frac{\hbar {k_F}^2}{m}$), (a) $\delta$-function barrier model ($H_a=0.1H_0$, $H_b=5\frac{\hbar {k_F}^2}{m}$), (b) finite thickness barrier model (see Fig. 3(c)) with a thin barrier ($H_a=0.1H_0$, $U=4\frac{\hbar {k_F}^2}{m}$, $d=\frac{1.18}{k_F}$), (c) finite thickness barrier model with a thick barrier ($H_a=0.1H_0$, $U=1.02\frac{\hbar {k_F}^2}{m}$, $d=\frac{5.51}{k_F}$).
  }
  \end{figure*}
\par
All these results are taken together, the shape of the superconductor junction, as well as the conditions of the insulating layer, have a large effect on the peak shift.
The large variety of the Doppler shift detected in past experiments is reasonably understood to reflect the difference in experimental details around the interface.
Therefore, in order to perform quantitative analysis on the Doppler shift, it is necessary to conduct spectroscopy by controlling the shape of the superconductor and barrier condition in detail.
The shape effect and vortex arrangement effect treated in the present analysis are expected to strongly affect the Volovik effect in specific heat measurements on polycrystals.
This effect will be analyzed in future works.

\section{Conclusion}
We have analyzed the origin of distributed experimental results on the Doppler shift in tunneling conductance spectra of $d$-wave superconductors.
The peak shifts due to the Doppler shift in tunneling spectroscopy are shown to respond sensitively to the shape of the superconductor and the barrier potential.
The peak shifts are enhanced by the demagnetizing effect at thin films, whereas the shifts are suppressed by the corrugations, such as slits and dips, at the interface. 
It was also found that the peak shift responds nonlinearly to the magnetic field due to the penetration of magnetic flux inside the superconductor. 
The variety in the shift amplitude reported in past experiments is consistent with the Doppler shift at realistic interface shapes. 
For future work, detailed control of the superconductor shapes and barrier conditions is required in order to achieve qualitative consistency in theory and experiments.
These results clearly demonstrate that tunnel spectroscopy has extremely high sensitivity to the interface conditions.
\section{Acknowledgments}
We acknowledge fruitful discussions with Y. Hiramatsu, K. Donguhn, R. Yano, and H. Taniguchi. 
This work was supported by JSPS KAKENHI Grant Numbers 23K17668, 24K00556, and 24K00583.

\bibliography{main}

\end{document}